\documentclass{article}

\usepackage{arxiv}

\usepackage[utf8]{inputenc} 
\usepackage[T1]{fontenc}    
\usepackage{hyperref}       
\usepackage{url}            
\usepackage{booktabs}       
\usepackage{amsfonts}       
\usepackage{nicefrac}       
\usepackage{microtype}      
\usepackage{fancyhdr}       
\usepackage{graphicx}       
\usepackage{authblk}        
\usepackage{amsmath}
\usepackage{multirow}

\usepackage[square,numbers]{natbib}
\title{Outlier Detection in Mendelian Randomisation
}

\author[1]{\large{\textbf{Maximilian M Mandl}\thanks{
Correspondence: Maximilian Mandl, Institute for Medical Information Processing, Biometry, and Epidemiology, Ludwig-Maximilians-Universität München, Germany. mmandl@ibe.med.uni-muenchen.de} }}
\author[1]{\textbf{Anne-Laure Boulesteix}}
\author[2,3]{\textbf{Stephen Burgess}}
\author[4,5,6]{\textbf{Verena Zuber}}

\affil[1]{\normalsize Institute for Medical Information Processing, Biometry, and Epidemiology, Faculty of Medicine, Ludwig-Maximilians-Universität, München, Germany}

\affil[2]{MRC Biostatistics Unit, School of Clinical Medicine, University of Cambridge, Cambridge, United Kingdom}

\affil[3]{Department of Public Health and Primary Care, British Heart Foundation Cardiovascular Epidemiology Unit, University of Cambridge, Cambridge, United Kingdom}

\affil[4]{Department of Epidemiology and Biostatistics, Imperial College London, London, United Kingdom}

\affil[5]{MRC Centre for Environment and Health, School of Public Health, Imperial College London, London, United Kingdom}

\affil[6]{Dementia Research Institute at Imperial College, Imperial College London, London, United Kingdom}

\begin{document}
\maketitle

\begin{abstract}
Mendelian Randomisation (MR) uses genetic variants as instrumental variables to infer causal effects of exposures on an outcome. One key assumption of MR is that the genetic variants used as instrumental variables are independent of the outcome conditional on the risk factor and unobserved confounders. Violations of this assumption, i.e. the effect of the instrumental variables on the outcome through a path other than the risk factor included in the model (which can be caused by pleiotropy), are common phenomena in human genetics. Genetic variants, which deviate from this assumption, appear as outliers to the MR model fit and can be detected by the general heterogeneity statistics proposed in the literature, which are known to suffer from overdispersion, i.e. too many genetic variants are declared as false outliers. We propose a method that corrects for overdispersion of the heterogeneity statistics in uni- and multivariable MR analysis by making use of the estimated inflation factor to correctly remove outlying instruments and therefore account for pleiotropic effects. Our method is applicable to summary-level data.
\end{abstract}


\newpage
\section{Introduction}\label{introduction}

Identification of causal effects in biomedical sciences is a challenging task. Most causal inference methods rely on specific assumptions which must be properly tested in practice. Mendelian Randomisation (MR) is an instrumental variable approach that uses genetic variants to infer causal effects of risk factors on an outcome \cite{davey2003mendelian}. Due to the randomisation of the genetic variants during meiosis, these can be used as instrumental variables that can potentially meet the restrictive methodological requirements naturally. Thus, causal effects can be consistently inferred even if unobserved confounders are present. For example, relevant clinical questions that have been addressed using MR include the investigations of the effect of blood lipids on coronary heart disease (CHD), age-related macular degeneration (AMD) or Alzheimer's disease \cite{richardson2020evaluating, zuber2021high, burgess2017mendelian}, and the effect of vitamin D levels on Multiple Sclerosis (MS) \cite{mokry2015vitamin}. The instrumental variable assumptions underlying MR require that the genetic variants are independent of the outcome conditional on the risk factor and unobserved confounders, also known as the exclusion restriction assumption. Violations of this exclusion restriction assumption, i.e. the effect of the instrumental variables on the outcome through a path other than the risk factor included in the model, can be caused by horizontal pleiotropy, which is a common phenomenon in human genetics \cite{solovieff2013pleiotropy}.

Genetic variants which deviate from this assumption appear as outliers in the MR model fit and can be detected by general heterogeneity statistics proposed in the literature \cite{bowden2018invited}. In MR analysis, these statistics are often inflated due to the heterogeneity of genetic variants exerting their downstream effects on the exposures of interest, mis-matches of allele frequencies when data is integrated from distinct samples or the variant specific heterogeneity estimates not being normally distributed as a ratio of two normal distributions does not follow a normal distribution. This excess heterogeneity may impede the detection of outlying instruments using the traditional methods and result in the removal of too many IVs which are not true outliers that impact the causal effect estimate and consequently the conclusions drawn from the MR analysis.

In this paper, we propose GC-Q, a simple method that corrects for overdispersion of the heterogeneity statistics in uni- and multivariable MR analysis by making use of the estimated inflation factor to correctly remove outlying instruments, therefore accounting for pleiotropic effects (Section \ref{sec2}). As we show in an extensive simulation study and analysis of real data examples, our proposed method is more conservative in detecting outliers as existing methods because it removes the minimum number of instruments necessary to retain unbiased effect estimates. Moreover, GC-Q leads to a reduction of the type I error in detecting outlying genetic variants used as instruments compared to the existing methods based on Cochran's Q. 
    
 Moreover, we provide a comprehensive review of different outlier detection methods in uni- and multivariable MR.  The code for the simulation study and the real data example in this paper are provided on GitHub for the purpose of reproducibility.\footnote{\textbf{GitHub:} \url{https://github.com/mmax-code/MR_outliers}}. In the recently introduced phases classification for methodological research \cite{heinze2024phases}, our contribution can be assigned to phase 2: it presents and demonstrates the use of a new method on real data, and provides first simulation results suggesting that it is useful in some cases and worth being further considered in phase 3 studies.

\section{Methods}\label{sec2}

In this section, we first give a brief overview of univariable und multivariable MR, and how horizontal pleiotropy violates the exclusion restriction assumption of instrumental variable analysis. Next, we discuss how heterogeneity statistics can be used to detect violations of this assumption and how specific pleiotropic genetic variants can be detected as outliers. We further show limitations of existing implementations of heterogeneity statistics and we introduce our novel method, GC-Q. Finally, we end with an overview and comparison of existing outlier detection methodologies for MR.

Regarding the notation, we examine the causal effect $\theta$ of a risk factor $X$ on an outcome $Y$ using genetic variants $G_i$ for $i = 1,...,n$ as instrumental variables (IVs). 
Subsequently, in a multivariable MR model we consider multiple causal effects $\theta_j$ ($j = 1,...,d$)  for multiple risk factors $X_j$ ($j = 1,...,d$) on an outcome $Y$. Following the most common MR design \cite{hartwig2016two}, real data examples are based on two-sample summary-level data to take advantage of large sample sizes and thus improve the precision of the estimates \cite{burgess2015using}. Additionally, all of our derivations are based on summary-level data. We therefore assume that the associations of genetic variants with the risk factor(s) and the outcome, and the causal effect of the risk factor(s) on the outcome, are linear and homogeneous. These assumptions have already been discussed in the literature \cite{burgess2016combining}.

\subsection{Univariable Mendelian Randomisation}

In order to define a \textit{valid} IV, the genetic variants in the univariable MR analysis require the following assumptions to hold \cite{greenland2000introduction}:

\begin{itemize}
    \itemsep0em 
    \item IV1(U):  Each  genetic variant $G_i$  for $i=1,...,n$   is associated with the exposure.
    \item IV2(U):  Each  genetic variant $G_i$  for $i=1,...,n$   is not associated with any confounder of the risk factor-outcome association. 
    \item IV3(U):  Each  genetic variant $G_i$ for $i=1,...,n$   is independent of the outcome $Y$ conditional on the risk factor $X$ and confounders $U$. 
\end{itemize}

\begin{figure*}
\centerline{\includegraphics[scale=0.2]{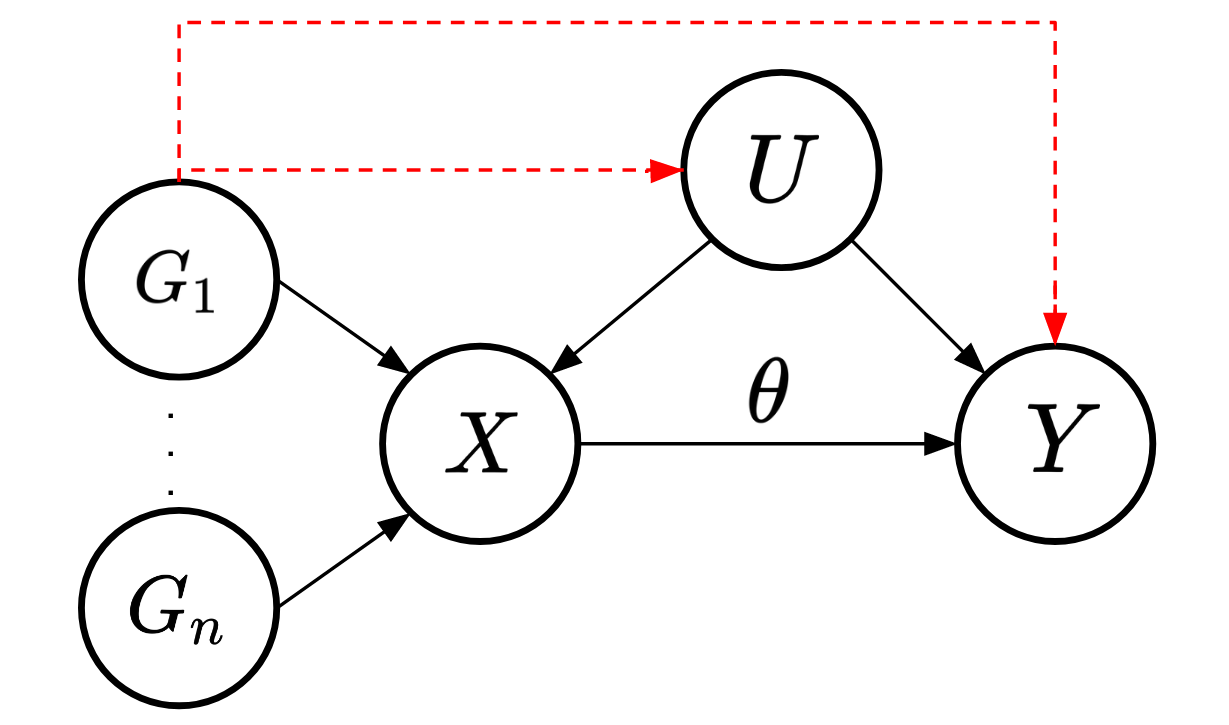}}
\caption{Causal directed acyclic graph (DAG) for the univariable Mendelian Randomisation setting. Genetic variants are denoted as $G_i$ for $i \in 1,...,n$, the set of confounders as $U$ and the causal effect of the risk factor $X$ on the outcome $Y$ being $\theta$. The red dashed lines represent the effect of the instrumental variable(s) on the outcome through paths other than the risk factor included in the model, e.g. caused by pleiotropy. \label{MR}}
\end{figure*}

Figure \ref{MR} shows the causal DAG for the univariable MR setting. Each genetic variant $G_i$ should only have an effect on the outcome via the risk factor. Pleiotropy is defined as the effect of any genetic variant $G_i$ that contains an effect via an independent pathway, i.e. not through the included risk factor in the MR model (red dashed lines in Figure \ref{MR}). Therefore, IV3 would be violated.

If IV1(U)--IV3(U)\footnote{Note that also linearity and homogeneity assumptions must hold.} hold, the consistent estimate of the causal effect $\theta$ is the inverse-variance weighted (IVW) estimate \cite{johnson2013efficient}

\begin{equation}
    \hat{\theta} = \frac{\sum_i^n \omega_i \hat{\theta_i}}
    {\sum_i^n \omega_i}, 
\end{equation}

\noindent where $\hat{\theta_i}$ is the ratio estimate of the $i$th IV, defined as $\hat{\theta_i} = \frac{\hat{\beta}_{Y_{i}}}{\hat{\beta}_{X_{i}}}$, where in the summary-level data setting $\hat{\beta}_{X_{i}}$ and $\hat{\beta}_{Y_{i}}$ are the genetic effects of IV $G_i$ on $X$ and $Y$ for variant $i$ respectively. The IV-specific inverse-variance weight $\omega_i$ is the precision of the respective ratio estimate. The estimate based on individual-level data can be obtained via the two-stage least-squares (2-SLS) approach \cite{angrist1995two}. The 2-SLS estimate is equivalent to the IVW estimate \cite{burgess2015integrating}.  However, in finite samples this is only true if all of the instruments are perfectly uncorrelated with each other. 

\subsection{Multivariable Mendelian Randomisation}

\begin{figure*}
\centerline{\includegraphics[scale=0.2]{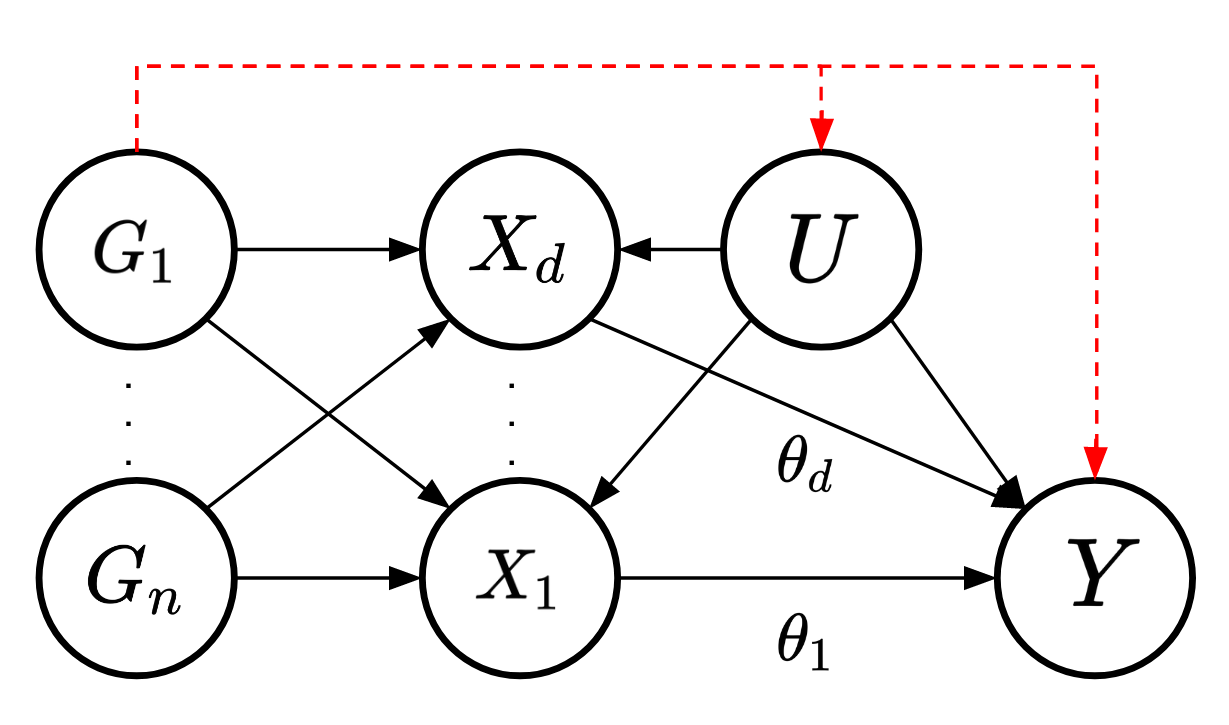}}
\caption{Causal directed acyclic graph (DAG) for the multivariable Mendelian Randomisation setting. Genetic variants $G_i$ ($i \in 1 ... n$), set of confounders $U$ and causal effects of the risk factors $X_j$ ($j \in 1 ... d$) on the outcome $Y$ being $\theta_j$. The red dashed lines represent the effect of the instrumental variable(s) on the outcome through paths other than the risk factors included in the model, e.g. caused by pleiotropy. \label{MVMR}}
\end{figure*}

As an extension to the standard MR approach, multivariable MR includes multiple potential risk factors in one joint model accounting for measured pleiotropy (see Figure \ref{MVMR}). In order to define a \textit{valid} IV, the genetic variants in the multivariable MR analysis require the following assumptions to hold  
for each genetic variant $G_i$ where $i=1,...,n$
\cite{burgess2013mendelian}:

\begin{itemize}
    \itemsep0em 
    \item IV1(M):  Each  genetic variant $G_i$  for $i=1,...,n$  is associated with at least one of the risk factors $X_j$.
    \item IV2(M):   Each genetic variant  $G_i$ for $i=1,...,n$  is not associated with any confounder of the risk factor-outcome associations. 
    \item IV3(M):  Each  genetic variant  $G_i$ for $i=1,...,n$  is independent of the outcome $Y$ conditional on the risk factors $X_j$  for $j = 1,...,d$  and confounders $U$.
\end{itemize}

Moreover, the following assumptions relate to which risk factors $X_j$ for $j = 1,...,d$ can be included in a multivariable MR model: 
\begin{itemize}
    \item RF1(M) Each risk factor $X_j$ for $j = 1,...,d$ needs to be strongly instrumented by at least one genetic variant $G_i$ for $i=1,...,n$, also denoted as relevance assumption.
    \item RF2(M) Each risk factor $X_j$ for $j = 1,...,d$ considered in the analysis cannot be linearly explained by the genetic associations of any other risk factor $X_j$ $for j = 1,...,d$ or by the combined genetic associations of several other risk factors included in the analysis, also denoted as no mulit-collinearity assumption.
\end{itemize}

If IV1(M)--IV3(M)\footnote{Note that also linearity and homogeneity assumptions must hold.} hold, the consistent estimates of the direct causal effects $\theta_j$ can be obtained from individual-level data via a 2-SLS approach or through the multivariable two-sample summary-level IVW method, with weights $se(\hat{\beta}_{Y_{i}})^{-2}$ being the inverse of the estimated variance for genetic variant $i$ \cite{burgess2015multivariable} and $\hat{\beta}_{X_{ij}}$, and $\hat{\beta}_{Y_{i}}$ being the genetic effects of $G_i$ on $X_{ij}$ and $Y_i$ for variant $i$ and risk factor $j$, respectively

\begin{equation}
    \hat{\beta}_{Y_{i}} = \sum_{j=1}^d \theta_j \hat{\beta}_{X_{ij}} + \varepsilon_i.
\end{equation}

\subsection{Heterogeneity Statistics}
\label{sec:hetero}

Tests for heterogeneity in the MR setting examine the null hypothesis that all genetic variants follow the same causal pathways through which the risk factors $X_1,...,X_d$ act on the outcome $Y$. The following heterogeneity statistics are based on Cochran's Q and compute weighted sums of squared residuals and differ in the variance factors they use for weighting \cite{greco2015detecting}. Cochran's Q was established in meta-analysis for the detection of heterogeneity between studies. Two-sample MR can be viewed as a meta-analysis over genetic variants used as IVs. Analogously, the sample size or the number of studies included equals the number of genetic variants used as IVs. Previous research has shown that the power of Cochran's Q increases with the number of studies and the total information available (total weight or inverse variance) and decreases substantially if a large proportion of the total information is based on one study \cite{hardy1998detecting}. On the other hand, the test arguably shows \lq\lq excessive'' power when the number of large studies increases \cite{higgins2003measuring}. Translating this into the MR framework means that the Q-statistic is likely to miss true heterogeneity (not rejecting the null hypothesis) when there are few genetic variants as IVs available and detects false heterogeneity (rejecting the null hypothesis) when there are many genetic variants available. Usually, genome-wide association studies (GWAS) entail high statistical power given their huge case numbers ($n>$ 1 million), i.e. we are more concerned about detecting too many SNPs as outliers. Therefore, we introduce a modification of the Q-statistic which allows the calibration of the standard Q-statistic, in order to reduce the \lq\lq excessive'' power of the test and decrease the type I error. 

Cochran’s Q-statistic was first applied as a global test to identify the presence of any invalid instruments in two-sample summary data MR with a single exposure by del Greco et al.\cite{greco2015detecting}. A generalised version of the Q-statistic for multivariable MR\cite{sanderson2019examination}
is defined as    

\begin{equation}
    Q = \sum_{i=1}^n \left(\frac{1}{\omega_i}\right) \left(\hat{\beta}_{Y_{i}}-\sum_{j}^d \hat{\theta}_{j} \hat{\beta}_{X_{ij}} \right)^2 \sim \chi^2_{(n-d)},
\end{equation}

\noindent with $i$ being the SNP index, $j$ being the risk factor index, and $\omega_i$ being the SNP-specific weight which can be approximated either using first or second order weights \cite{bowden2018improving}. Under the null hypothesis, Cochran's Q follows a $\chi^2_{n-d}$ distribution with $n-d$ degrees of freedom. First order weights are simply defined as  $\omega_i = \sigma^2_{Y_i}$ with $\sigma_{Y_i}$ being the standard error of $\hat{\beta}_{Y_i}$. The first order weights are an approximation relying on the so-called no measurement error (NOME) assumption which assumes that the standard errors of the exposures associations are negligible \cite{bowden2018improving}. These first-order weights are known to lead to an overdispersion in the heterogeneity statistic resulting in an inflation of the type 1 error rate, that is detecting heterogeneity when it is not present \cite{bowden2018improving}.

In applied analysis, using the two-sample summary-level MR setting, there are additional sources of excess heterogeneity due to:
\begin{itemize}
\itemsep0em
\item Wide-spread but negligible pleiotropic effects
\item Small disagreements in allele frequencies between the first sample used to derive the exposure association and the second independent sample used to derive the outcome associations
\item The variant-specific estimates being not normally distributed (as the ratio of two normal distributions is not normal)
\end{itemize}

Bowden et al \cite{bowden2018improving} and  Sanderson et al. \cite{sanderson2019examination} propose an adjusted weighting scheme of the Cochran Q-statistic to test for invalid instruments in the two-sample univariable \cite{bowden2018improving} and multivariable \cite{sanderson2019examination} summary-level data setting using the following modified second order weights based on a Taylor expansion of the ratio estimate. In contrast to the first order weights that are proportional to the uncertainty of the standard error of the genetic associations with the outcome, the second order weights also account for the standard error of the genetic associations with the exposure and thus model uncertainty in both the nominator and denominator of the ratio estimate. 

The second order weights for multivariable MR are defined as
$\omega_i = \sigma^2_{Y_i} + \sum_{j}^d \hat{\theta}_{j}^2 \sigma_{X_{ij}}^2 + \sum_{jk} 2 \hat{\theta}_{j} \hat{\theta}_{k} \sigma_{X_{ijk}}$, with $i$ being the SNP index, $j$ being the risk factor index, $\sigma_{Y_i}$ and $\sigma_{X_{ij}}$ being the standard error of $\hat{\beta}_{Y_i}$ and $\hat{\beta}_{X_{ij}}$ respectively, and $\sigma_{X_{ijk}}$ being the covariance for all pairs $\hat{\beta}_{X_{ij}}$ and $\hat{\beta}_{X_{ik}}$ for $j,k \in 1,...,d$ and $j \neq k$. Importantly, the covariance term $\sigma_{X_{ijk}}$ for exposures $j$ and $k$, which is necessary in multivariable MR cannot be estimated from the data at hand and can only be calculated from individual-level data, which is often not available in practice \cite{sanderson2019examination}.

\subsection{Outlier detection in MR}

The main aim of outlier detection in MR is the detection of invalid IVs with strong pleiotropic effects that can when included into the MR model bias the causal effect estimate $\hat{\theta}$ and consequently distort conclusions drawn from the MR analysis. The objective is to identify these individual genetic instruments with pleiotropic effects which appear as outliers to the MR model fit. The local test statistic $q_i$ of SNP $i$ defined as

\begin{equation}
    q_i = \left(\frac{1}{\omega_i}\right) \left(\hat{\beta}_{Y_{i}}-\sum_{j}^d \hat{\theta}_{j} \hat{\beta}_{X_{ij}} \right)^2
\end{equation}

has been proposed for outlier detection \cite{bowden2018invited}. 
Under the null hypothesis, $q_i$ asymptotically follows a $\chi^2_{(1)}$ distribution with one degree of freedom. Note, that we must correct for multiple testing if we test individual genetic instruments, for example, using a Bonferroni correction, i.e. dividing the significance level by the number of instruments $n$ \cite{bowden2018improving}. A conservative multiple testing procedure is recommended in order to only remove clear outliers and to retain as many genetic variants as possible as instrumental variables. Yet, outlier detection using the current implementation of the local $q-$statistic is impeded  by overinflation when using first-order weights. In contrast, when using second-order weights in multivariable MR, one essential parameter (covariance term $\sigma_{X_{ijk}}$) is not readily available when working with summary-level data and consequently often set to zero in practice.

\subsection{Correction for Overdispersion with Genomic Control (GC-Q)}

We suggest correcting for overdispersion of the first-order weighted heterogeneity statistics in MR-analysis by making use of the estimated inflation factor to remove outlying instruments which may be invalid due to horizontal pleiotropy. The idea of correcting for overdispersion is based on the Genomic Control approach which was originally used in the context of GWAS \cite{devlin1999genomic}. More precisely, the Genomic Control approach was developed for testing if a large set of genomic markers or SNPs are associated with a quantitative trait of interest. Typically, when performing genome-wide testing of genetic markers, like in GWAS, only a small proportion of genetic markers are associated with a trait of interest and the large majority of genetic markers can be considered as following the null model. Yet, Devlin and Roeder observed that even these null genetic markers do not follow the theoretical null distribution of the statistical association test, but display overdispersion which is constant across the genome \cite{devlin1999genomic}. In GWAS, this observed overdispersion is due to population stratification, cryptic relatedness, or unobserved confounding \cite{devlin2001genomic}.

In analogy with genetic association tests as described in Devlin and Roeder \cite{devlin1999genomic}, the local heterogeneity statistic $q_i$ of instrument $i$ follows a $\chi^2$ distribution with one degree of freedom and a non-centrality parameter $0$ under the Null ($\chi_1^2(0)$), i.e. 

\begin{equation}
q_i/\lambda \stackrel{H_0}{\sim} \chi_1^2(0),
\end{equation}

\noindent where $\lambda$ is the overdispersion parameter and constant for all SNPs. Thus the empirical distribution of the local $q_i$-statistic is inflated from $\chi_1^2(0)$ to $\lambda \chi_1^2(0)$. This means that no heterogeneity is present or in other words, the instrument is valid. This follows from the general Cochran's statistic in Section~\ref{sec:hetero}, that is generally inflated. \\
From a Bayesian perspective the distribution of the local $q-$statistic -- with outlying SNPs being present -- can be modeled using a mixture model of two $\chi^2$ distributions, where the distribution under the Null ($\chi_1^2(0)$) is representing the valid IVs and $\chi_1^2(A_i^2)$ is the distribution with non-centrality parameter $A_i^2 > 0$ associated with the i--th outlier

\begin{equation}
q_i/\lambda \stackrel{H_1}{\sim} \rho \chi_1^2(A_i^2) + (1 - \rho) \chi_1^2(0),
\end{equation}

\noindent where $\rho$ is the prior probability that a given SNP is an outlier as indicated by excess heterogeneity and consequently invalid. As Devlin and Roeder\cite{devlin1999genomic} propose, a simple frequentist estimate of the inflation parameter $\hat{\lambda}$ can be derived from the data as

\begin{equation}\label{lambda}
\hat{\lambda} = \frac{\tilde{q}}{0.675^2},
\end{equation}

\noindent with $\tilde{q}$ being the median of $q_i$ for all $i=1,...,n$ SNPs and $0.675^2$ being the median of the theoretical $\chi_1^2$ distribution.

An important assumption when estimating  $\hat{\lambda}$ according to \eqref{lambda} is that at least half of the genetic variants used as IVs are valid instruments. This assumption is common to the median MR approach \cite{bowden2016consistent} and more general for any type of outlier detection approach \cite{rousseeuw2011robust}. 
For all $i=1,...,n$ SNPs, we reject the Null, i.e. SNP $i$ is considered as outlier, if $q_i/\lambda > \chi_{1,\alpha^*}^2$ with $\chi_{1,\alpha^*}^2$ being the critical value at level $\alpha^*$ and $\alpha^* = \alpha/n$ being the Bonferroni adjusted significance level to provide a conservative multiple testing adjustment.    

The local heterogeneity statistic $q^{adj}_i$ can subsequently formulated as
\begin{equation}
      q^{adj}_i = \left(\frac{1}{\lambda}\right)\left(\frac{1}{\omega_i} \right) \left(\hat{\beta}_{Y_{i}}-\sum_{j}^d \hat{\theta}_{j} \hat{\beta}_{X_{ij}} \right)^2. 
\end{equation}

\subsection{Other methods to detect outliers in summary-level MR}

\subsubsection{Heterogeneity statistics with second order weights}

Sanderson et al. \cite{sanderson2019examination} propose an adjustment to Cochran's Q in the two-sample summary setting for testing the presence of horizontal pleiotropy as the standard version of Cochran's Q merely has a weighting of the variance of $\beta_{Yi}$ denoted as $\sigma^2_{Yi}$, and is thus not asymptotically $\chi^2$ distributed. Therefore, they make use of second-order weights, 

\begin{equation}
    \omega_i = \sigma^2_{Yi} + \sum_{j}^d \hat{\beta}_{X_{ij}} \sigma^2_{X_{ij}} + \sum_{j}^d \sum_{\substack{k \\ j \neq k}}^d \sigma_{ijk}
\end{equation}

where $\hat{\beta}_{X_{ij}}$ are efficient estimators of the causal effects and $\sigma_{ijk}$ are the covariances of the exposures which need to be estimated from individual-level data.

\subsubsection{MR--PRESSO}

Verbanck et al. \cite{verbanck2018detection} developed the MR-PRESSO method to detect pleiotropy (global test), the correction for pleiotropy via outlier removal (outlier test), and test for significant distortions in the causal estimates before and after the outlier removal (distortion test).  The MR-PRESSO global test is defined as the following residual sum of squares (RSS)
\begin{equation}
    RSS = \sum_{i=1}^n \left(\frac{1}{\omega_i}\right) \left(\hat{\beta}_{Y_{i}}-\sum_{j} \hat{\theta}_{j}^{-i} \hat{\beta}_{X_{ij}} \right)^2 \sim \chi^2_{(n-d)}
\end{equation}
where  $\omega_i$ are the first-order weights and
$\hat{\theta}_{j}^{-i}$ is the causal effect estimate from an IVW MR model without variant $i$. The respective $p-$values are calculated using a simulation procedure. The main difference between the RSS of MR--PRESSO and the $Q-$statistic is that the causal effect estimate of MR--PRESSO is calculated excluding the $i$th IV and that $p-$values are derived in a non-parametric fashion using a simulation procedure which scales with the number of IVs and becomes prohibitively slow when including hundreds of IVs. 

As with the local q-statistics, the outlier test aims at detecting individual SNPs as outliers. For a given genetic variant $i$, the observed RSS defined as $\left(\frac{1}{\omega_i}\right) \left(\hat{\beta}_{Y_{i}}-\sum_{j} \hat{\theta}_{j}^{-i} \hat{\beta}_{X_{ij}} \right)^2$ is compared to the distribution of the expected simulated residual sum of squares. The detection mechanism can be described as follows: For each variant the causal effect $\hat{\theta}_{-i}$ is computed without variant $i$. Afterwards, the observed residual sum of squares is compared to the expected residual sum of squares. Finally, an empirical $p$-value that is Bonferroni adjusted is computed to decide whether variant $i$ is an outlier or not.

\subsubsection{Radial MR}

The Galbraith Radial plot, adapted for MR, plots the $z$-statistics for genetic variant $i$, which is the ratio estimate $\hat{\theta}_i$ divided by its standard error, against the precision of the ratio estimate which is equal the inverse standard error \cite{bowden2018radial}. This is particularly relevant when different IVs have varying precision and consequently contribute with different weights to the final causal estimate. Moreover, the Radial MR approach allows for a flexible use of first or second order weights and can adapt an intercept in the MR model fit which is also known as the MR-Egger approach \cite{bowden2015egger}. Outlier detection is based on the heterogeneity statistics as described above.

\subsubsection{Summary and comparison of outlier detection methodologies for MR}

To conclude the Methods section, we present an overview of outlier detection methodologies for MR.
The $Q$-statistic with first order weights is easy to estimate, but relies on the NOME assumption. Consequently, it is not well calibrated and shows an overinflation. In contrast, the Q-statistic with second order weights is well calibrated, but needs additional parameters which cannot be estimated from summary-level data alone. Moreover, since the second order weights include the causal effect estimate and iterative estimation procedure needs to be implemented \cite{bowden2018improving}. MR-PRESSO relies on a permutation procedure which is computationally expensive when the number of genetic variants used as IVs increases. 

Our newly proposed GC-Q approach is a recalibrated version of the first-order weights Q-statistic, which can be estimated from the data at hand and does not rely on computationally intensive permutation procedures. As we are going to show in an extensive simulation study in the next Section, GC-Q selects the minimum number of potential outliers necessary to achieve unbiased MR effect estimates. All outlier detection methods perform a Bonferroni correction for multiple testing of the individual IV heterogeneity statistics with the aim to be as conservative as possible and to only remove the minimum number of outliers necessary to avoid any bias of the MR model.

\section{Simulation Study}\label{sec3}

The primary objective of this simulation study is to compare the performance of different methods to detect outlying genetic variants used as IVs, i.e. SNPs that entail pleiotropic effects and violate the exclusion restriction assumption. For this purpose, we mainly consider a scenario reflecting directional pleiotropy, i.e. some of the genetic variants $G$ are consistently positively associated with the outcome $Y$ through a different causal pathway than the risk factors for both uni- and multivariable MR. However, we show results for the balanced pleiotropy setting in Table \ref{tabuni_balanced}. \\

The simulation study is set up as follows\footnote{The simulation setup is similar for the univariate case with only one risk factor.} for each individual $x \in 1, ..., 500,000$, we simulate $100$ genetic variants from a binomial distribution where the minor allele frequency is a probability that is drawn uniformly between [0.01, 0.5]. In the following, three $\beta$-coefficients are simulated from a normal distribution with $\beta_{X_j} \sim \textrm{N(1,2)}$ for the first-stage regression, with $j \in 1,...,d$ being the index of risk factors. We fix the variance explained for the first stage regression at 15\% for all risk factors and the confounder. The calculated variances are used to simulate correlated error terms for the first-stage regression from a multivariate normal distribution, i.e. $\epsilon_i \sim \textrm{MVN}(\mu,\Sigma)$ with $\mu = (0,0,0)$ and $\Sigma$ being a positive-definite covariance matrix to simulate a medium correlation between the risk factors. The outlying SNPs are simulated as an additional unknown risk factor with a coefficient $\rho$ which is equal to zero for ($1-p$)\% of the SNPs and drawn from a uniform or normal distribution for the remaining $p$\%. The variance explained for the second-stage regression is fixed at 50\% with causal effects set to 0, 1, and -0.5, for the three risk factors and to 1 for the additional variable that represents the unmeasured pleiotropic pathway that creates outliers (i.e. the unobserved risk factor). From the individual-level data, summary-level data on genetic associations is generated and the different methods are compared for different measures, namely the sensitivity and specificity of the outlier classification,  the mean bias of the causal effect estimates ($\frac{1}{z}\sum_{i=1}^{z} \hat{\theta_i} - \theta_i$), the mean squared error ($\frac{1}{z}\sum_{i=1}^{z} (\hat{\theta_i} - \theta_i)^2$), the average number of detected outliers in relation to the true outlier rate, and the average absolute number of detected outliers. In total $z = 1,000$  simulation runs were performed for each setting. Note that the parameter settings are inspired by Sanderson et al.\cite{sanderson2019examination}. The code for the simulation and real data analysis is available on GitHub\footnote{\textbf{GitHub:} \url{https://github.com/mmax-code/MR_outliers}}. \\

The competing methods are referred to as follows: \textit{Full model} denotes the estimated model with all SNPs, \textit{Standard} denotes the standard Cochran's $q$-statistic based on first-order weights, \textit{Sanderson} denotes the adjusted $q$-statistic by Sanderson et al. based on second-order weights \cite{sanderson2019examination}, \textit{MR-Presso} refers to the outlier test by Verbanck et al.\cite{verbanck2018detection},
\textit{MR-Radial} describes the MR-Radial method using modified second order weights\cite{bowden2018radial}, and \textit{GC-Q} refers to the newly proposed method based on the calibrated first-order weights.

\begin{figure}[hbt!]
\begin{center}
\includegraphics[scale=0.7]{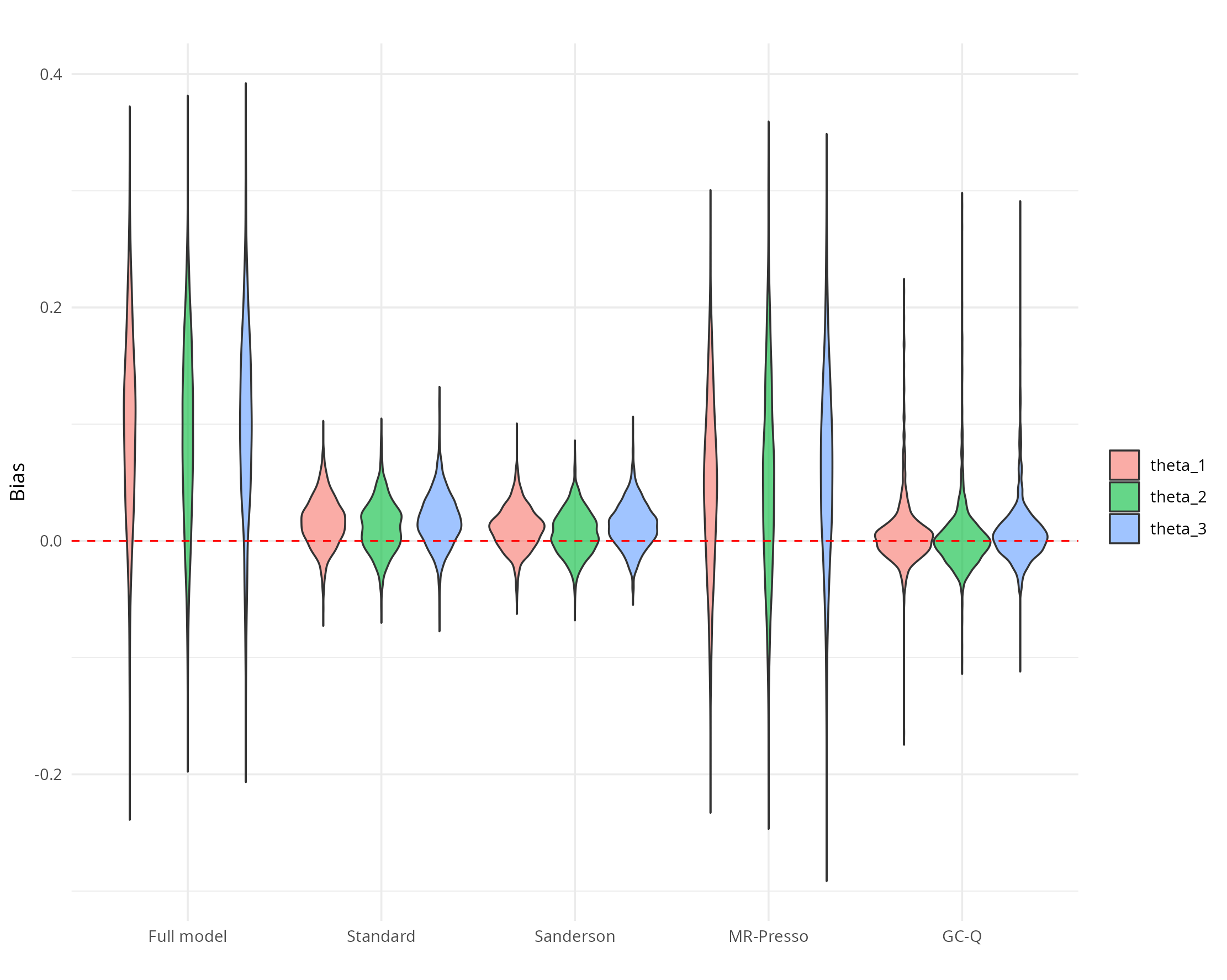}
\caption{Violin plots for the bias of the causal effect estimates of $\theta_1$ (red), $\theta_2$ (green), and $\theta_3$ (blue) in the multivariate simulation setting (15\% outliers) after outlier adjustment and for the full model.}
\label{violin}
\end{center}
\end{figure}

As Table \ref{tabuni} and \ref{tabmulti} show, GC-Q outperforms the other methods in terms of specificity (true negative rate) with nearly 100\% for the uni- and multivariable simulation settings. The specificity is similar for MR-Presso and the GC-Q version in the multivariate setting. With regard to the sensitivity (true positive rate), the GC-Q method performs as well as the other methods for most univariable settings and nearly as well as the Standard and Sanderson methods for the multivariable setting (except for the 20\% outliers case). Note that the GC-Q method performs still better than MR-Presso (57\%). With an increasing outlier rate, the sensitivity of the GC-Q method decreases. This is due to an increasing bias of the median of the q-statistics and expected since by design GC-Q works up to 25\% outlying SNPs for directional pleiotropy and analogously up to 50\% for the balanced case. 

This means that our method is more conservative in terms of detecting outlying SNPs than the Standard and Sanderson approaches, i.e. we can observe less power but a smaller type I error. The average number of detected outliers is always closer to 1 for GC-Q. This illustrates the conservative behaviour of GC-Q to only remove the minimum number of outliers necessary to obtain an unbiased causal effect estimate. Our method outperforms the other methods with regard to the bias for most settings, showing that the smaller number of outliers removed by GC-Q also provides the causal effect estimate which is closest to the actual simulated effect. Even though this is true for a small number of outliers, the Sanderson method should be preferred if more than 25\% outlying SNPs are expected. The violin plots in Figure \ref{violin} show that even though the GC-Q method exhibits a higher variance than the Standard and Sanderson methods, on average the bias is centered close to zero in contrast to the other methods that have a positive bias. The advantage of GC-Q becomes obvious with stronger outlier effects. As Table \ref{tabuni_strong} shows, GC-Q still performs well in terms of bias, MSE, and average number of detected outliers while the results of the other methods seem to be more harmed by pleiotropic effects that are stronger. In addition, the behavior of GC-Q is as expected if balanced pleiotropy occurs. Table \ref{tabuni_balanced} once again depicts its conservatism in terms of sensitivity and its features with regard to bias and MSE.

\section{Real Data Application}\label{sec4}

In this section, we compare the results of different heterogeneity measures for uni- and multivariable MR based on real data with regard to Vitamin D as exposure for Multiple Sclerosis and blood lipids as candidate exposures for coronary heart disease, age-related macular degeneration, and Alzheimer's.

\subsection{Univariable MR: Vitamin D as exposure for Multiple Sclerosis}

The following application example considers circulating vitamin D levels as exposures for multiple sclerosis (MS) in the univariable MR setting. Summary-level data on genetic associations with vitamin D are derived from $361,194$ individuals and taken from UK Biobank\footnote{\url{https://biobank.ctsu.ox.ac.uk/crystal/field.cgi?id=30890}}. Summary-level data on genetic associations, with the outcome MS including $14,498$ European ancestry cases and $24,091$ European ancestry controls, was taken from the International Multiple Sclerosis Genetics Consortium (IMSGC)\footnote{\url{https://www.ebi.ac.uk/gwas/studies/GCST005531}}\cite{beecham2013analysis}.

As instrumental variables we selected $n=22$ independent (clumping threshold of $r^2 < 0.001$) genetic variants associated with vitamin D at genome-wide significance ($p-$value $< 5 \times 10^{-8}$). As shown in Table \ref{tabunidata} the GC-Q $q$-values did not detect any outlier, while the Standard and Sanderson's adjusted $q$-values detected the same single outlier, rs4944958. MR-Presso and MR-Radial detected two outliers, rs4944958 and rs7041. Genetic variants associated with vitamin D are known to belong to three tiers, the direct vitamin D pathway, U/V absorption and the cholesterol metabolism\cite{manousaki2020genome}. Of interest, the one outlier identified by MR-Presso, MR-Radial, the Standard and Sanderson $q$-values, rs4944958, is an intron of the \textit{NADSYN1} gene which affects a precursor of cholesterol and is part of cholesterol metabolism which may indicate horizontal pleiotropy. In contrast, rs7041, which was only identified by MR-Presso and MR-Radial, is located in the \textit{GC} gene, also known as vitamin D-binding protein, which is part of the direct vitamin D pathway and unlikely to reflect other biological pathways and may represent a false positive finding.

Interestingly, the effect size of the MR analysis depends on the outlier removal approach, with the MR-Presso and MR-Radial approach that removed more genetic variants having the strongest protective effect estimate as can be seen in Table \ref{tabunidata}.

\subsection{Multivariable MR: Blood lipids as candidate exposures}

As a second application example, we consider blood lipids as exposures in a multivariable MR setting following Burgess and Davey Smith\cite{burgess2017mendelian} who analysed if genetically predicted levels of low-density lipoprotein cholesterol (LDL-C), high-density lipoprotein cholesterol (HDL-C), and triglycerides are associated with age-related macular degeneration (AMD) and used coronary heart disease (CHD) as a positive control where clear evidence for lipids being a causal risk factor for CHD exists\cite{holmes2015mendelian}. In contrast, evidence for lipids being causal for Alzheimer's disease is mixed and not supported by MR studies\cite{larsson2017modifiable}. Independent genetic variants were selected as IVs if they were associated with any of the blood lipids at genome-wide significance \cite{burgess2017mendelian} resulting in $n=185$ IVs. 

Table \ref{tabmultidatasnp} shows the genetic variants used as IVs and detected as outliers by the different approaches. In general, the GC-Q approach detected the fewest outliers, followed by MR-Presso, which is in line with the simulation results for the methods based on Cochran's Q. The Standard $q$-value approach and Sanderson's $q$-values detected exactly the same outliers. The difference in detection of outliers is reflected in the respective MR estimates as shown in Table \ref{tabmultidataeffect}. For CHD, all methods yield similar effect sizes independent of the removal for outliers. In contrast, for the outcomes AMD and Alzheimer's disease, the MR effect estimates not only differ in their effect sizes but also in their significance. For example, the effect estimate of HDL-cholesterol on AMD is significant at the 5\% level for the full model without outlier removal, whereas it doubles its effect size and is even significant at the 0.001 level after outlier removal with the other methods. 
The benefit of outlier removal is most striking for the effect of LDL-cholesterol on Alzheimer's disease. Here, including all genetic variants associated with any blood lipid, provided significant evidence for genetically predicted levels of LDL-cholesterol to be associated with Alzheimer's disease. The removal of outliers, in particular of one genetic variant, rs6859, in the \textit{APOE} gene region, leads to an insignificant MR effect estimate.

\section{Related work}

Some researchers prefer to use robust methods, such as the MR-Median approach to avoid dealing with outliers. But firstly, we are more interested in detecting outlying SNPs and thus robust estimation methods such as the Median-MR approach would not be helpful in this endeavor. And secondly, these approaches have their own specific disadvantages, e.g., MR-Median has less power compared to other methods. If we use the MR-median method in our real data application in section~\ref{sec4} on the effect of vitamin D on multiple sclerosis, we get an estimate of $-0.296$ (0.164) with a p-value of $0.07$. Even though the Standard and Sanderson method perform similarly in the effect size, we end up with higher standard errors and a change of significance of the causal estimate. Nonetheless, these methods offer an alternative to circumvent the issue of outliers in MR analyses with respect to less biased causal effect estimates. Table~\ref{tab_robust} we benchmarked GC-Q with the Median-based method \cite{bowden2016consistent} and the MR-RAPS method \cite{mrraps}. For the setting with 10\% outliers, all methods perform similarly well. However, the bias and MSE get very large in the setting with 80\% outliers for MR-RAPS and GC-Q, with an advantage for the Median-based method with respect to the bias and the MSE. GC-Q performs similiarly to the full model without any outliers removed.

\section{Limitations}

An important aspect of introducing new methods is to highlight their respective limitations. As we have already mentioned, GC-Q works for up to 25\% outlying SNPs for directional pleiotropy and analogously up to 50\% for the balanced case. Table~\ref{tab50_80} shows its performance in extreme cases with 50\% and 80\% of the SNPs being outliers in the univariate MR setting for directional pleiotropy, i.e. the assumptions of the methods do not hold anymore.
 
As the estimation of $\hat{\lambda}$ is dependent on the median of the unadjusted local $q$-statistics, we observe an extremely inflated distribution over the simulation runs. $\hat{\lambda}$ is on average inflated by a factor of 100, which results in a deflation of the local q-statistics and subsequently no outlying SNPs can be found. GC-Q thus performs similar to the full model without any outliers removed. 

As Table~\ref{tab50_80} the other outlier detection methods still work better in the 50\% setting -- the bias, however, is not negligible and seems to gradually approach the bias of the full model with higher outlier proportions. Since $\hat{\lambda}$ of the GC-Q method can no longer be estimated correctly and is based on the median of the $\chi^2_1$ distribution, there might be a way to adjust it with a simple correction factor for the hyper-inflated median of the unadjusted q-statistic for settings where its assumptions do not hold. As a reliable rule of thumb for the correction factor is beyond the scope of this manuscript, we leave it open for future investigations.

\section{Discussion}\label{sec5}

Overdispersion in the heterogeneity statistic is a common problem in meta-analysis \cite{higgins2003measuring, hardy1998detecting} and is not limited to MR. For example, in meta-analysis of clinical trials, heterogeneity may arise because of a diverse range of factors including diversity in doses, lengths of follow up, study quality, and inclusion criteria for participants \cite{higgins2003measuring}. In MR, heterogeneity can be caused by different molecular pathways affecting the exposure. For example, there genetic variation acts via many different biological pathways on obesity including the metabolism, cholesterol transport, fat storage, appetite regulation, food preference, reward mechanisms and physical exercise. The heterogeneity test statistic is known to depend on the sample size\cite{hardy1998detecting} which is the number of studies included in a meta-analysis or in MR, the number of genetic variants used as IVs. The power is low when there are few SNPs; in contrast, the heterogeneity statistic shows substantial overdispersion when there are many SNPs. Powerful GWAS have identified hundreds of regions in the genome associated with potential exposures and provide a large number of IVs making the calibration of the heterogeneity statistic an important statistical problem.

Here, we propose GC-Q an adjusted version of the local q-statistic to detect outliers. GC-Q has the potential to decrease the type I error at the price of a reduced power (see Table \ref{tabuni} and \ref{tabmulti}). With this method, we correct for overdispersion of the heterogeneity statistics by making use of the estimated inflation factor using a mixture model approach. 

GC-Q is using the first-order weights, which in contrast to the second-order weights do not include the precision of the genetic association with the exposure, which is also known as the \textit{No Measurement Error} (NoME) assumption. 
Another important assumption of GC-Q is that less than half of the IVs are invalid (in the balanced pleiotropy setting), an assumption that is necessary to guarantee the identifiability of the mixture model. In order to estimate the over-dispersion parameter GC-Q requires a minimum number of IVs and is only recommended for polygenic exposures where there are many genetic variants available as IVs. The mixture model on which GC-Q is formulated has been shown in simulations to perform well on 50 observations and conservative when less observations are available \cite{BACANU20001933}.
In addition, GC-Q performs especially well if the outlier effect is strong (see Table \ref{tabuni_strong}).
Another advantage of GC-Q is that it does not require additional parameters as the observational covariance between exposures, and it does not require a two-step procedure for estimation (note the second-order weights require the causal effect estimate and can only be obtained in an iterative procedure). MR-PRESSO uses a computationally expensive simulation procedure to define the Null-distribution which becomes computationally more expensive as the number of instruments grows. In contrast, GC-Q is based on a fast and simple computational implementation which uses first-order weights and relies on a closed-form mixture model formulation. Thus, no iterative procedure to calculate the weights or simulations are needed to define the Null distribution. A disadvantage of GC-Q is that it relies on the assumption that less than half of the IVs are valid, but this assumption is common to all other methods which rely on outlier detection, including MR-PRESSO, Radial MR, or the Median MR method. 

However, we do not claim that our method outperforms the existing methods in all cases. We see our approach as complementary to the outlier detection methods in MR analysis. 

When removing outliers in MR models it is necessary to strike a balance between removing all invalid IVs that may bias the causal effect estimate and retaining the largest number of IVs to retain the largest sample size possible.

As we show in our simulation study and in the real data analysis GC-Q removes the smallest number of outliers while obtaining unbiased causal effect estimates, which highlights that GC-Q is conservative and removes only the minimum number of invalid IVs necessary to obtain the unbiased causal effect.

Let us finish by a recent quote from Strobl and Leisch \cite{strobl2024against}. They claim that \lq\lq the research question 'What is the best method in general' is ill-posed'' and warn methodological researchers who present new methods against the \lq\lq one method fits them all'' philosophy. In this spirit, we emphasize that our method certainly cannot be recommended universally for all datasets and in all contexts (especially in an early phase paper such as ours \cite{heinze2024phases}), but shows a promising behavior in practically relevant situations. With this in mind, we do not want to claim that our newly introduced method outperforms existing methods in every analytical setting and can be seen as complementary within the MR literature.  

\section*{Acknowledgments}

This study was partly funded by DFG grant BO3139/7-1 to Anne-Laure Boulesteix. Verena Zuber is supported by the United Kingdom Research and Innovation Medical Research Council grant MR/W029790/1. All data used in our study is in the public domain. Our study is based on publicly available summary-level data on genetic associations from the International AMD Genetics Consortium \url{http://amdgenetics.org/}, the GWAS catalog \url{https://www.ebi.ac.uk/gwas/}, MAGNETIC NMR-GWAS \url{http://www.computationalmedicine.fi/data}, the International Multiple Sclerosis Genetics Consortium \url{https://imsgc.net/}, and UK-Biobank \url{http://www.nealelab.is/uk-biobank}.

\subsection*{Author contributions}

MM and VZ designed the study. MM implemented the method, performed the simulations, analyzed the data and interpreted the results. MM and VZ prepared the initial manuscript draft. VZ directed the project. MM, ALB, VZ, and SB reviewed and edited the manuscript.

\subsection*{Conflict of interest}

The authors have declared no conflict of interest. 

\subsection*{Data Availability Statement}

The data and code can be found on \textbf{GitHub:} \url{https://github.com/mmax-code/MR_outliers}.

\newpage
\bibliography{references}

\newpage

\begin{center}
\begin{table}[hbt!]%
\centering
\caption{Simulation results for outlier detection in the univariable MR scenario. Sensitivity and specificity for detecting outliers. Mean bias and MSE for the causal effect estimate $\hat{\theta}$, average number of detected outliers in relation to the true outlier rate ($\overline{p}$), and average number of detected outliers ($\overline{a}$).\label{tabuni}}%
\begin{tabular*}{500pt}{@{\extracolsep\fill}lcccccc@{\extracolsep\fill}}
\toprule
\textbf{Measure} & \textbf{Full model}  & \textbf{Standard}  & \textbf{Sanderson}  & \textbf{MR-Presso} & \textbf{MR-Radial} &  \textbf{GC-Q} \\
\midrule


 \multicolumn{7}{c}{5\% outliers} \\
 \midrule
Sensitivity & -  & 1.00  & 1.00 & 1.00 & 1.00 & 0.99\\
Specificity & -  & 0.97  & 0.99  & 0.99 & 0.90 & 1.0\\
Mean Bias & 0.041  & 0.002  & -0.001  & 0.0001 & -0.0001 & -0.002  \\
MSE & 0.004 & 0.0002 & 0.0001 & 0.004 & 0.0005 & 0.0001\\
$\overline{p}$ & - & 1.40  & 1.11  & 1.18  & 2.81 & 1.01 \\
$\overline{a}$ & -  &   7.01   & 5.55     & 5.88  &   14.04 &  5.05  \\

\midrule
 \multicolumn{7}{c}{10\% outliers} \\
 \midrule
Sensitivity & -  & 1.00  & 1.00 & 1.00 & 1.00 & 0.99\\
Specificity & -  & 0.94  & 0.97  & 0.97 & 0.84 & 1.0\\
Mean Bias & 0.09  & 0.01  & 0.003  & 0.005 & 0.03 & -0.002  \\
MSE & 0.01 & 0.0004 & 0.0002 & 0.01 & 0.001 & 0.0001\\
$\overline{p}$ & - & 1.55  & 1.23  & 1.29  & 2.47 & 0.99 \\
$\overline{a}$ & -  &   15.45    & 12.27   &  12.89     &  24.72 & 9.93 \\

\midrule
 \multicolumn{7}{c}{15\% outliers} \\
 \midrule
Sensitivity & -  & 1.00  & 1.00 & 1.00 & 1.00 & 0.97\\
Specificity & -  & 0.88  & 0.94  & 0.93 & 0.76 & 1.0\\
Mean Bias & 0.13  & 0.02  & 0.01  & 0.01 & 0.04 & 0.0007  \\
MSE & 0.02 & 0.001 & 0.0003 & 0.02 & 0.003 & 0.0002\\
$\overline{p}$ & - & 1.66  & 1.34  & 1.39  & 2.37 & 0.97 \\
$\overline{a}$ & -  &   24.92    & 20.06   &  20.88     &  35.54 & 14.52    \\

\midrule
 \multicolumn{7}{c}{20\% outliers} \\
 \midrule
Sensitivity & -  & 1.00  & 1.00 & 1.00 & 1.00 & 0.91\\
Specificity & -  & 0.82  & 0.89  & 0.88 & 0.68 & 1.0\\
Mean Bias & 0.18  & 0.03  & 0.02  & 0.02 & 0.06 & 0.02  \\
MSE & 0.04 & 0.002 & 0.0006 & 0.04 & 0.005 & 0.002\\
$\overline{p}$ & - & 1.73  & 1.42  & 1.47  & 2.26 & 0.91 \\
$\overline{a}$ & -  &   34.60    &  28.38    &  29.40  &  45.20 & 18.23   \\
\bottomrule

\end{tabular*}
\end{table}
\end{center}
\null
\vfill

\newpage

\begin{center}
\begin{table}[hbt!]%
\centering
\caption{Simulation results for outlier detection in the multivariable MR scenario. Sensitivity and specificity for detecting outliers. Mean bias and MSE for the causal effect estimates $\hat{\theta}_1$, $\hat{\theta}_2$, and $\hat{\theta}_3$, average number of detected outliers in relation to the true outlier rate ($\overline{p}$), and average number of detected outliers ($\overline{a}$).\label{tabmulti}}%
\begin{tabular*}{500pt}{@{\extracolsep\fill}lccccc@{\extracolsep\fill}}
\toprule
\textbf{Measure} & \textbf{Full model}  & \textbf{Standard}  & \textbf{Sanderson}  & \textbf{MR-Presso}  &  \textbf{GC-Q} \\
\midrule


 \multicolumn{6}{c}{5\% outliers} \\
 \midrule
Sensitivity & -  & 1.00  & 1.00 & 0.71 & 1.00 \\
Specificity & -  & 0.98  & 0.99  & 1.00 & 1.00 \\
Mean Bias & 0.03  & 0.002  & 0.0005  & 0.01 & -0.0004   \\
MSE & 0.003 & 0.0002 & 0.0002 & 0.002 & 0.0002 \\
$\overline{p}$ & - & 1.31  & 1.12  & 0.71  & 1.01 \\
$\overline{a}$ & -  &  6.57    & 5.61     &  3.57     &  5.04 \\


\midrule
 \multicolumn{6}{c}{10\% outliers} \\
 \midrule
Sensitivity & -  & 1.00  & 1.00 & 0.66 & 1.00 \\
Specificity & -  & 0.94  & 0.96  & 1.00 & 1.00 \\
Mean Bias & 0.06  & 0.007  & 0.004  & 0.03 & -0.0004  \\
MSE & 0.008 & 0.0004 & 0.0003 & 0.004 & 0.0002 \\
$\overline{p}$ & - & 1.58  & 1.32  & 0.66  & 1.00 \\
$\overline{a}$ & -  &  15.82    & 13.17     &  6.65     &  10.01 \\


\midrule
 \multicolumn{6}{c}{15\% outliers} \\
 \midrule
Sensitivity & -  & 1.00  & 1.00 & 0.62 & 0.98 \\
Specificity & -  & 0.86  & 0.91  & 1.00 & 1.00 \\
Mean Bias & 0.09  & 0.02  & 0.01  & 0.05 & 0.006  \\
MSE & 0.02 & 0.0007 & 0.0005 & 0.009 & 0.0008 \\
$\overline{p}$ & - & 1.81  & 1.53  & 0.62  & 0.98  \\
$\overline{a}$ & -  &   27.14    & 22.92   &  9.28     &  14.65    \\

\midrule
 \multicolumn{6}{c}{20\% outliers} \\
 \midrule
Sensitivity & -  & 1.00  & 1.00 & 0.57 & 0.77 \\
Specificity & -  & 0.77  & 0.83  & 1.00 & 1.00 \\
Mean Bias & 0.13  & 0.02  & 0.02  & 0.09 & 0.06  \\
MSE & 0.03 & 0.001 & 0.0008 & 0.02 & 0.01 \\
$\overline{p}$ & - & 1.92  & 1.67  & 0.57  & 0.78  \\
$\overline{a}$ & -  &   38.46    &  33.38    &  11.46  & 15.62   \\
\bottomrule

\end{tabular*}
\end{table}
\end{center}
\null
\vfill

\newpage

\begin{center}
\begin{table}[hbt!]%
\centering
\caption{Simulation results for outlier detection in the univariable MR scenario with a strong outlier effect ($p=15\%$). Sensitivity and specificity for detecting outliers. Mean bias and MSE for the causal effect estimate $\hat{\theta}$, average number of detected outliers in relation to the true outlier rate ($\overline{p}$), and average number of detected outliers ($\overline{a}$).\label{tabuni_strong}}%
\begin{tabular*}{500pt}{@{\extracolsep\fill}lcccccc@{\extracolsep\fill}}
\toprule
\textbf{Measure} & \textbf{Full model}  & \textbf{Standard}  & \textbf{Sanderson}  & \textbf{MR-Presso} & \textbf{MR-Radial} &  \textbf{GC-Q} \\
\midrule

Sensitivity & -  & 1.00  & 1.00 & 1.00 & 1.00 & 0.98\\
Specificity & -  & 0.60  & 0.67  & 0.66 & 0.47 & 1.0\\
Mean Bias & 0.54  & 0.11  & 0.07  & 0.07 & 0.19 & 0.007 \\
MSE & 0.40 & 0.02 & 0.01 & 0.40 & 0.05 & 0.003\\
$\overline{p}$ & - & 3.25  & 2.85  & 2.90  & 4.02 & 0.98 \\
$\overline{a}$ & -  &   48.70   & 42.69     & 43.50  &   60.37 &  14.68  \\
\bottomrule

\end{tabular*}
\end{table}
\end{center}

\newpage
\begin{center}
\begin{table}[hbt!]%
\centering
\caption{Simulation results for outlier detection in the univariable MR scenario with balanced pleiotropy ($p=20\%$). Sensitivity and specificity for detecting outliers. Mean bias and MSE for the causal effect estimate $\hat{\theta}$, average number of detected outliers in relation to the true outlier rate ($\overline{p}$), and average number of detected outliers ($\overline{a}$).\label{tabuni_balanced}}%
\begin{tabular*}{500pt}{@{\extracolsep\fill}lcccccc@{\extracolsep\fill}}
\toprule
\textbf{Measure} & \textbf{Full model}  & \textbf{Standard}  & \textbf{Sanderson}  & \textbf{MR-Presso} & \textbf{MR-Radial} &  \textbf{GC-Q} \\
\midrule

Sensitivity & -  & 1.00  & 1.00 & 1.00 & 1.00 & 0.99\\
Specificity & -  & 0.96  & 0.98  & 0.97 & 0.86 & 1.0\\
Mean Bias & -0.0005  & -0.002  & -0.003  & -0.003 & -0.003 & -0.002 \\
MSE & 0.009 & 0.0004 & 0.0002 & 0.009 & 0.001 & 0.0003\\
$\overline{p}$ & - & 1.18  & 1.09  & 1.11  & 1.57 & 0.99 \\
$\overline{a}$ & -  &   23.59   & 21.78     & 22.24  &  31.37 &  19.73  \\
\bottomrule
\end{tabular*}
\end{table}
\end{center}
\null
\vfill

\newpage

\begin{center}
\begin{table}[hbt!]%
\centering
\caption{Real data analysis: Causal effect estimates, standard errors, p-values, and outlying SNPs for vitamin D on multiple sclerosis for the univariable MR scenario.\label{tabunidata}}%
\begin{tabular*}{500pt}{@{\extracolsep\fill}lcccc@{\extracolsep\fill}}
\toprule
\textbf{Method}  & \textbf{Causal estimate $\hat{\theta}$} & \textbf{Std. Error} & \textbf{P-Value} & \textbf{SNP}  \\
\midrule
Full model & $-0.44$ & 0.10 & 0.0003 & - \\
Standard & $-0.30$  & 0.09 & 0.0048 & rs4944958  \\
Sanderson & $-0.30$ & 0.09 & 0.0048  & rs4944958   \\
MR-Presso        & $-0.53$ &  0.12  & 0.0002  & rs4944958, rs7041 \\
MR-Radial        & $-0.54$ &  0.10  & 0.00001  & rs4944958, rs7041 \\
GC-Q   & $-0.44$ & 0.10  & 0.0003 & -  \\
\bottomrule
\end{tabular*}
\end{table}
\end{center}

\newpage

\begin{center}
\begin{table}[hbt!]%
\centering
\caption{Real data analysis: Outlying SNPs in the multivariate MR scenario. Data from Burgess and Davey Smith\cite{burgess2017mendelian}, originally from the Global Lipids Genetics Consortium \cite{willer2013discovery} and Fritsche et al.\cite{fritsche2016large}.\label{tabmultidatasnp}}
  \begin{tabular}{p{4.5cm}  p{3.3cm}  p{3.1cm}  p{3.5cm}}
\toprule
\textbf{Method}  &  \textbf{CHD} & \textbf{AMD} & \textbf{ALZ} \\
\midrule
    Standard & rs1250229, rs4530754, rs579459, rs12801636, rs653178, rs6489818, rs952044  & rs1883025, rs653178, rs1532085, rs261342  rs9989419, rs6859, rs492602 &  rs1883025, rs17788930, rs6859  \\
\midrule
    Sanderson & rs1250229, rs4530754, rs579459, rs12801636, rs653178, rs6489818, rs952044  & rs1883025, rs653178, rs1532085, rs261342  rs9989419, rs6859, rs492602 &  rs1883025, rs17788930, rs6859  \\
\midrule
    MR-Presso    & rs4530754, rs12801636, rs653178, rs952044      & rs1883025, rs1532085, rs261342, rs9989419, rs6859  & rs17788930, rs6859 \\
\midrule
    GC-Q   & rs653178      & rs1532085, rs261342 & rs6859  \\
\bottomrule
\end{tabular}
\end{table}
\end{center}

\newpage

\begin{center}
\begin{table}[hbt!]%
\centering
 \caption{Real data analysis: Causal effect estimates for LDL- ($\hat{\theta_1}$), HDL-cholesterol ($\hat{\theta_2}$), and triglycerides ($\hat{\theta_3}$) on coronary heart disease, macular degeneration, and Alzheimer's. Data from Burgess et al.\cite{burgess2017mendelian}\label{tabmultidataeffect}.}
\begin{tabular}{p{2cm}  p{1.3cm} p{1.3cm} p{1.5cm}  p{1.3cm} p{1.3cm} p{1.5cm}  p{1.3cm} p{1.3cm} p{1.5cm}}
\toprule
\multirow{2}{*}{\textbf{Method}} &
      \multicolumn{3}{c}{\textbf{CHD}} &
      \multicolumn{3}{c}{\textbf{AMD}} &
      \multicolumn{3}{c}{\textbf{ALZ}} \\
       & Estimate & Std. Err. & P-Value & Estimate & Std. Err. & P-Value & Estimate & Std. Err. & P-Value  \\
\midrule 
    Full Model \\
    $\hat{\theta_1}$ & 0.39  & 0.04 & $2\times 10^{-16}$ & -0.04 & 0.07 & 0.55 & 0.24 & 0.08 & 0.002 \\
    $\hat{\theta_2}$ & -0.07 & 0.05 & 0.116 & 0.18 & 0.08 & 0.028 & -0.11 & 0.09 & 0.206 \\
    $\hat{\theta_2}$ & 0.14  & 0.06 & 0.012 & -0.07 & 0.10 & 0.437 & -0.14 & 0.11 & 0.186 \\
\midrule 
    Standard \\
    $\hat{\theta_1}$ & 0.41   & 0.04 & $2\times 10^{-16}$  & -0.03 & 0.06 & 0.646    & 0.09 & 0.05 & 0.055 \\
    $\hat{\theta_2}$ & -0.06  & 0.04 & 0.165        &  0.35 & 0.08 & $1.7\times 10^{-5}$ &-0.06 & 0.05 & 0.238  \\
    $\hat{\theta_3}$ & 0.14   & 0.05 & 0.005        & 0.10 & 0.08 & 0.250        & -0.09 & 0.06 & 0.161 \\
\midrule 
    Sanderson \\
    $\hat{\theta_1}$ & 0.41   & 0.04 & $2\times 10^{-16}$ & -0.03 & 0.06 &  0.646    &0.09 & 0.05 & 0.055 \\
    $\hat{\theta_2}$ & -0.06 & 0.04 & 0.165       & 0.35 & 0.08 & $1.7\times 10^{-5}$ & -0.06 & 0.05 & 0.238  \\
    $\hat{\theta_3}$ & 0.14   & 0.05 & 0.005       & 0.10 & 0.08 & 0.250       &-0.09 & 0.06 & 0.161 \\
\midrule 
    MR-Presso \\
    $\hat{\theta_1}$ & 0.41   & 0.04  & $2\times 10^{-16}$ & -0.03 & 0.06 & 0.640     &0.08 & 0.05 & 0.082 \\
    $\hat{\theta_2}$ & -0.05 & 0.04  & 0.201       & 0.34 & 0.08 & $1.7\times 10^{-5}$ & -0.10 & 0.05 & 0.074  \\
    $\hat{\theta_2}$ & 0.13   & 0.05  & 0.010       & 0.14 & 0.08 & 0.106     & -0.12 & 0.07 & 0.078 \\
\midrule 
    GC-Q \\
    $\hat{\theta_1}$ & 0.40  & 0.04 & $2 \times 10^{-16}$   & -0.06 & 0.06 & 0.324     & 0.09 & 0.05 & 0.081 \\
    $\hat{\theta_2}$ & -0.06 & 0.05 & 0.155         & 0.47 & 0.08 & $1.7\times 10^{-5}$ & -0.08 & 0.06 & 0.148  \\
    $\hat{\theta_3}$ & 0.14  & 0.05 & 0.011         & 0.18 & 0.09 & 0.045     & -0.11& 0.07 & 0.091 \\
\bottomrule 
\end{tabular}
\end{table}
\end{center}
\null
\vfill

\newpage

\begin{center}
\begin{table}[hbt!]%
\centering
\caption{Simulation results for outlier detection in the univariable MR scenario. Mean bias and MSE for the causal effect estimates $\hat{\theta}$ for GC-Q, MR-RAPS (with overdispersion and L2 loss), and the Median-based method. For details on the parameter settings see Section~\ref{sec3}. \label{tab_robust}}%
\begin{tabular*}{500pt}{@{\extracolsep\fill}lcccc@{\extracolsep\fill}}
\toprule
\textbf{Measure} & \textbf{Full model}  & \textbf{GC-Q}  & \textbf{Median}  & \textbf{MR-RAPS}  \\
\midrule

 \multicolumn{5}{c}{10\% outliers} \\
 \midrule
Mean Bias & 0.087  & -0.002  & 0.002  & 0.090   \\
MSE & 0.011  & 0.0001  & 0.0002  & 0.011  \\

\midrule
 \multicolumn{5}{c}{80\% outliers} \\
 \midrule
Mean Bias & 0.72  & 0.65  & 0.39  & 0.72   \\
MSE  & 0.54  & 0.50  & 0.26  & 0.54  \\

\bottomrule

\end{tabular*}
\end{table}
\end{center}
\null
\vfill

\newpage

\begin{center}
\begin{table}[hbt!]%
\centering
\caption{Simulation results for outlier detection in the univariable MR scenario. Sensitivity and specificity for detecting outliers. Mean bias and MSE for the causal effect estimates $\hat{\theta}$, average number of detected outliers in relation to the true outlier rate ($\overline{p}$), and average number of detected outliers ($\overline{a}$).\label{tab50_80}}%
\begin{tabular*}{500pt}{@{\extracolsep\fill}lcccccc@{\extracolsep\fill}}
\toprule
\textbf{Measure} & \textbf{Full model}  & \textbf{Standard}  & \textbf{Sanderson}  & \textbf{MR-Presso} & \textbf{MR-Radial} &  \textbf{GC-Q} \\
\midrule

 \multicolumn{7}{c}{50\% outliers} \\
 \midrule
Sensitivity & -  & 0.99  & 0.99 & 0.99 & 1.00 & 0.01\\
Specificity & -  & 0.54  & 0.65  & 0.64 & 0.42 & 1.0\\
Mean Bias & 0.45  & 0.14  & 0.13  & 0.12 & 0.18 & 0.45  \\
MSE & 0.22 & 0.04 & 0.05 & 0.22 & 0.05 & 0.22\\
$\overline{p}$ & - & 1.45  & 1.33  & 1.35  & 1.57 & - \\
$\overline{a}$ & -  &   72.67   & 66.74   & 67.30  &   78.70 &  -  \\

\midrule
 \multicolumn{7}{c}{80\% outliers} \\
 \midrule
Sensitivity & -  & 0.96  & 0.94 & 0.94 & 0.97 & 0\\
Specificity & -  & 0.39  & 0.50  & 0.49 & 0.31 & 1.0\\
Mean Bias & 0.72  & 0.63  & 0.67  & 0.66 & 0.60 & 0.72  \\
MSE & 0.54 & 0.49 & 0.54 & 0.54 & 0.46 & 0.54\\
$\overline{p}$ & - & 1.12  & 1.06  & 1.07  & 1.15 & - \\
$\overline{a}$ & -  &   89.25    & 85.14   &  85.47     &  91.67 & - \\

\bottomrule

\end{tabular*}
\end{table}
\end{center}
\null
\vfill

\end{document}